\RequirePackage{fix-cm}

\documentclass[twocolumn,epjc3]{svjour3}  
\smartqed
\RequirePackage{graphics}
\RequirePackage{graphicx}
\RequirePackage{subfig}
\RequirePackage{multirow}
\RequirePackage{wrapfig}
\RequirePackage{fullpage}
\RequirePackage[switch]{lineno} 
\RequirePackage{array}
\RequirePackage{preprintcover}
\RequirePackage{xspace} 
\RequirePackage{atlasphysics} 
\RequirePackage[hyperindex,breaklinks]{hyperref} 

\newcommand{\pTphi}{p_{\mathrm{T,\phi}}}
\newcommand{\pTK}{p_{\mathrm{T},K}}
\newcommand{\pK}{p_{K}}
\newcommand{\yphi}{|y_{\phi}|}
\newcommand{\phiKK}{\phi \rightarrow K^{+}K^{-}}

\newcommand{\dedx}{\mathrm{d}E/\mathrm{d}x}
\newcommand{\PYS}{PYTHIA\kern +0.1em 6}
\newcommand{\PYE}{PYTHIA\kern +0.1em 8}

\PreprintCoverPaperTitle{The differential production cross section of the $\phi$(1020) meson in $\sqrt{s}=7$~\TeV\ pp collisions measured with the ATLAS detector}  
\PreprintIdNumber{CERN-PH-EP-2012-269}  
\PreprintCoverAbstract{A measurement is presented of the $\phiKK$ production cross section at $\sqrt{s}$ = 7~\TeV\ using $pp$ collision data corresponding to an integrated luminosity of 383~$\mathrm{\mu b^{-1}}$, collected with the ATLAS experiment at the LHC. Selection of $\phi$(1020) mesons is based on the identification of charged kaons by their energy loss in the pixel detector. The differential cross section is measured as a function of the transverse momentum, $\pTphi$, and rapidity, $y_{\phi}$, of the $\phi$(1020) meson in the fiducial region 500 $<\pTphi<$ 1200~\MeV, $|y_{\phi}|<$ 0.8, kaon $\pTK>$ 230~\MeV\ and kaon momentum $\pK<$ 800~\MeV. The integrated $\phi(1020)$-meson production cross section in this fiducial range is measured to be $\sigma_{\phiKK}$ = 570 $\pm$ 8 (stat) $\pm$ 66 (syst) $\pm$ 20 (lumi) $\mathrm{\mu b}$.}  
\PreprintJournalName{EPJC}

\journalname{Eur. Phys. J. C}
\begin{document}

\title{The differential production cross section of the $\phi$(1020) meson in
  $\sqrt{s}$ = 7~\TeV\ $pp$ collisions measured with the ATLAS Detector}

\author{The ATLAS Collaboration}     
\institute{CERN}

\date{Received: March 31 2014}
\maketitle
\begin{abstract}
A measurement is presented of the $\phiKK$ production cross section at $\sqrt{s}$ = 7~\TeV\ using $pp$ collision data corresponding to an integrated luminosity of 383~$\mathrm{\mu b^{-1}}$, collected with the ATLAS experiment at the LHC. Selection of $\phi$(1020) mesons is based on the identification of charged kaons by their energy loss in the pixel detector. The differential cross section is measured as a function of the transverse momentum, $\pTphi$, and rapidity, $y_{\phi}$, of the $\phi$(1020) meson in the fiducial region 500 $<\pTphi<$ 1200~\MeV, $|y_{\phi}|<$ 0.8, kaon $\pTK>$ 230~\MeV\ and kaon momentum $\pK<$ 800~\MeV. The integrated $\phi(1020)$-meson production cross section in this fiducial range is measured to be $\sigma_{\phiKK}$ = 570 $\pm$ 8 (stat) $\pm$ 66 (syst) $\pm$ 20 (lumi) $\mathrm{\mu b}$.  
\PACS{13.85.Ni \and 14.40.Df \and 14.40.Aq \and 14.40.Be}
\end{abstract}

\section{Introduction}\label{sec:intro}
Perturbative quantum chromodynamics (QCD) successfully describes physics at high momentum transfer, while phenomenological models are needed for soft interactions at lower momentum transfers. An accurate description of these soft interactions is required to model so-called “underlying events” present in hard scattering events.~Measurements of the $\phi$ (1020)-meson probe strangeness production at a soft scale $Q \sim$ 1~\GeV, which is sensitive to $s$-quark and low-$x$ gluon densities. Production of $\phi$(1020) mesons is also sensitive to fragmentation details and thus $\phi$(1020) measurements can constrain phenomenological soft hadroproduction models.

This paper presents a measurement with the \mbox{ATLAS} detector~\cite{Aad:2008zzm} of the $\phi(1020)$-meson production cross section in $pp$ interactions  at $\sqrt{s}=7$~\TeV, using the $\phi \rightarrow K^{+}K^{-}$ decay mode. The cross section is measured in bins of transverse momentum, $\pTphi$, or of rapidity $\yphi$.\footnote{ATLAS uses a right-handed coordinate system with its origin at the nominal interaction point (IP) in the centre of the detector and the $z$-axis along the beam pipe. The $x$-axis points from the IP to the centre of the LHC ring, and the $y$-axis points upward. Cylindrical coordinates ($r,\phi$) are used in the transverse plane, $\phi$ being the azimuthal angle around the beam pipe. The pseudorapidity is defined in terms of the polar angle $\theta$ with respect to the beamline as $\eta = - \mathrm{ln[tan}(\theta/2)]$.} The selection of $\phi$(1020)-meson candidates requires the identification of kaons in order to reduce the large combinatorial background from other charged particles. Charged particles are reconstructed with the inner detector, which consists of a silicon pixel detector, a microstrip semiconductor tracker (SCT), and a straw-tube transition radiation tracker (TRT). The inner detector barrel (end-cap) parts consist of 3 (2 $\times$ 3) pixel layers, 4 (2 $\times$ 9) layers of double-sided silicon strip modules, and 73 (2 $\times$ 160) layers of TRT straws. A track traversing the barrel typically has 11 silicon hits (3 pixel clusters, and 8 strip clusters), and more than 30 straw-tube hits.
The whole inner detector is immersed in a 2\,T axial magnetic field. The specific energy loss of charged particles in the pixel detector is used to identify low-momentum pions, kaons and protons~\cite{ATLAS-CONF-2011-016}.

To avoid model-dependent extrapolations outside the detector acceptance, the cross section is measured in the fiducial region, defined as 500 $<\pTphi<$ 1200~\MeV, $|y_{\phi}|<$ 0.8, kaon transverse momentum $\pTK>$ 230~\MeV\ and kaon momentum $\pK<$ 800~\MeV. In the region 0.8 $<\yphi<$ 1.0, $\phi$(1020) decays would only be accepted up to $\pTphi \sim$ 700~\MeV, because the requirement of $\pK<$ 800~\MeV\ has a lower efficiency at higher rapidity. The fiducial range is limited to the region where the differential cross section can be measured and where correcting for the losses due to the requirements on kaon momentum is reliable. The measurement is corrected for detector effects and can be compared directly with Monte Carlo (MC) generators at particle level.

Many measurements of the $\phi$(1020) production cross section have been performed at different centre-of-mass energies, using different decay modes and in different rapidity ranges. Among these are a study at $\sqrt{s}=7$~\TeV\ by \mbox{ALICE}~\cite{Abelev:2012hy} in a similar rapidity region and another by LHCb~\cite{Aaij2011267} in the forward rapidity region. The $\phi$(1020) production cross section presented in this paper is compared to the measurement by ALICE and to MC predictions.

\section{Data set and event selection}\label{sec:event-sel}
A data sample with an integrated luminosity of 383~$\mathrm{\mu  b^{-1}}$ from $pp$ collision data taken in April 2010 at $\sqrt{s}=7$~\TeV\  is used. The contribution of pile-up, i.e. multiple collisions per bunch crossing, is negligible for this data sample, with a peak luminosity of $1.8 \cdot 10^{28} \mathrm{cm^{-2}s^{-1}}$. The luminosity is measured in dedicated van der Meer scans with an estimated uncertainty of 3.5\%~\cite{Aad:2013ucp}. The data sample was selected with the minimum bias trigger scintillators (MBTS)~\cite{2010ir} to minimize any possible bias in the measured cross section. The MBTS are mounted at each end of the tracking detector in front of the liquid-argon endcap-calorimeter cryostats at $z$ = $\pm$ 3.56~m and were configured to require one hit above threshold from either side of the detector. This trigger is shown to be highly efficient in selecting inelastic $pp$ collisions~\cite{2010ir}. Tracks are fitted with a kaon-mass assumption to account for energy losses in the detector material. Events are required to contain at least two tracks with $\pT >$~150~\MeV\ and to have a primary vertex (PV, defined as the vertex in the event with the largest $\Sigma \pT$ over all reconstructed tracks associated to the vertex)~\cite{ATLAS-CONF-2010-069} reconstructed using the beam spot information~\cite{2010ir}.

MC simulations are used to correct the data for detector effects and to compare with the fully corrected data. The MC generators used are \texttt{\PYS} \cite{Sjostrand:2006za}, \texttt{\PYE}~\cite{Sjöstrand2008852}, \texttt{Herwig++}~\cite{Bahr:2008pv} and \texttt{EPOS} \cite{Werner:2005jf,Werner:2010aa}. Different versions of the same MC generator, that differ in sets of tunable parameters used in modeling the soft component of proton-proton interactions, are called tunes. Both \texttt{\PYS} and \texttt{\PYE} are general purpose generators which implement the Lund string hadronisation model~\cite{Andersson:1983ia} and describe non-diffractive interactions (including Multiple Parton Interactions, MPI) via lowest-order perturbative QCD, with phenomenological regularisation of the divergence of the cross section as $\pT \rightarrow 0$. Diffractive processes are included which involve the exchange of a colour singlet. Both inelastic non-diffractive and diffractive processes are mixed in accordance with the generator cross sections. The \texttt{PYTHIA} tunes considered are MC09~\cite{ATL-PHYS-PUB-2010-002} with \texttt{\PYS} version 6.421, DW~\cite{Albrow:2006rt} and Perugia0~\cite{Skands:2010ak} with \texttt{\PYS} version 6.423, and two A2 tunes with \texttt{\PYE} version 8.153, i.e. with the MSTW2008LO \cite{Martin:2009iq,springerlink:10.1007/JHEP03(2011)032} and CTEQ6L1 \cite{Pumplin:2002vw} PDF sets. The MC09 and Perugia0 tunes use a $\pT$-order- ed parton shower model with MPI and the initial-state shower interleaved in a common sequence of decreasing $\pT$. For the \texttt{\PYE} A2 tunes, the final-state showers are also interleaved in this way. The DW tune utilises the older virtuality-ordered parton shower which is not interleaved with MPI.

\texttt{Herwig++} version 2.5.1 is used with the UE7-2~\cite{Gieseke:2011xy} tune. \texttt{Herwig++} is also a general purpose generator but differs from \texttt{PYTHIA} in that it uses a cluster hadronisation model~\cite{Webber:1983if} and an angular-ordered parton shower. \texttt{Herwig++} contains a tunable eikonalised MPI model which assumes independence between separate scatters in the event. In order to simulate inelastic minimum bias events the following mechanism is used. For a fixed impact parameter, Poisson distributions are sampled to provide the number of soft and perturbatively-treated semi-hard scatters to simulate per event.

\texttt{EPOS} 1.99 v2965 is used with the \texttt{EPOS-LHC}~\cite{Pierog:2013ria} tune. \texttt{EPOS} contains a parametrised approximation of the hydrodynamic evolution of initial states using a parton based Gribov-Regge~\cite{Drescher:2000ha} theory which has been tuned to LHC data. 

The ATLAS detector is simulated~\cite{ATLASSimEPJC} using \texttt{GEANT4} \cite{Agostinelli2003250}. The reconstruction of $K^{\pm}$ tracks from $\phiKK$ decays generated by \texttt{\PYS} MC09 is used for the calculation of the tracking efficiency. A consistency test of the full $\phi$(1020)-meson reconstruction is performed with \texttt{\PYS} MC09 and \texttt{Herwig++} UE7-2.

As the $\phi$(1020) meson has no measurable decay length, only tracks originating from the PV are used. Each track must pass the following requirements: more than one pixel cluster and more than one SCT hit; $\pT>$ 230~\MeV; $p<$ 800~\MeV\ and $|\eta|<$ 2.0. The condition $\pT>$ 230~\MeV\ is adopted since the tracking efficiency for kaon tracks with $\pTK<$ 230~\MeV\ and central $|\eta|$ is close to zero. Kaons produced with such low momenta effectively deposit all their energy in the detector and support materials before reaching the SCT. The cut on track momentum of $p<$ 800~\MeV\ is dictated by particle identification requirements and is explained in the next section. 

\begin{figure}
  \includegraphics[width=8cm]{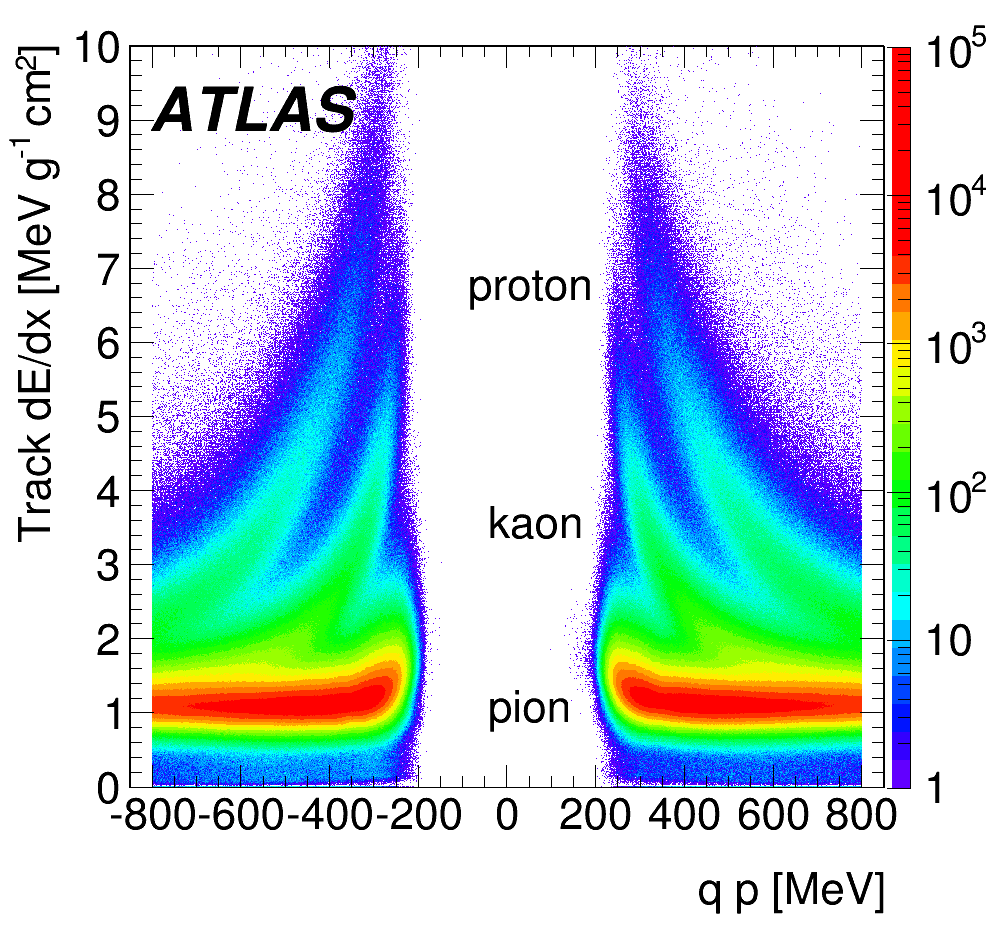}
\caption{The truncated mean (see text for detailed explanation) for the energy loss per track as a function of signed momentum for tracks accepted in the analysis. The bands corresponding to the energy lost by pions, kaons and protons are labelled.}
\label{fig:dEdx-vs-mom}
\end{figure}

\section{Particle identification}\label{sec:pid}
Every pair of oppositely charged tracks passing the tracking cuts is examined. The identification of a pair of tracks candidate for a $\phi \rightarrow K^{+}K^{-}$ decay requires a particle identification (PID) step to remove the large combinatorial background from pairs containing one or two charged particles that are not kaons. Discrimination between background (consisting mostly of pions) and kaons is achieved using energy loss in the pixel detector. The mean energy deposited by a charged particle is described by the Bethe--Bloch formula as a function of the particle's velocity~\cite{bethe-bloch}. For momenta larger than 1~\GeV, the energy lost by the particles starts to be dominated by relativistic effects and can no longer be used for particle identification. The mean energy loss per unit length is estimated as the energy deposited by a particle in the traversed layers of the pixel detector divided by the local thickness traversed in the detector material. The energy deposited is calculated after removing the pixel cluster with the largest charge for particles with three or four associated pixel clusters or after removing the two clusters with the largest charge for particles with more than four pixel clusters. The track $\dedx$ is calculated using a truncated mean of the $\dedx$~values of the individual pixel clusters as this gives a better~resolution than the simple mean. The expected energy loss for a kaon with $\pK=$ 500~\MeV\ is 2.4~\MeV\ $\mathrm{g^{-1}}$~$\mathrm{cm^{2}}$. For a pion with the same momentum, an energy loss of 1.2~\MeV\ $\mathrm{g^{-1}}$ $\mathrm{cm^{2}}$ is expected. The average energy loss per track as a function of signed momentum, $qp$, where $q$ is the particle charge, is shown in Fig.~\ref{fig:dEdx-vs-mom}; bands indicating pions, kaons and protons are clearly visible.

\begin{figure*}
  \centerline{
   \subfloat[]{\includegraphics[scale=0.25]{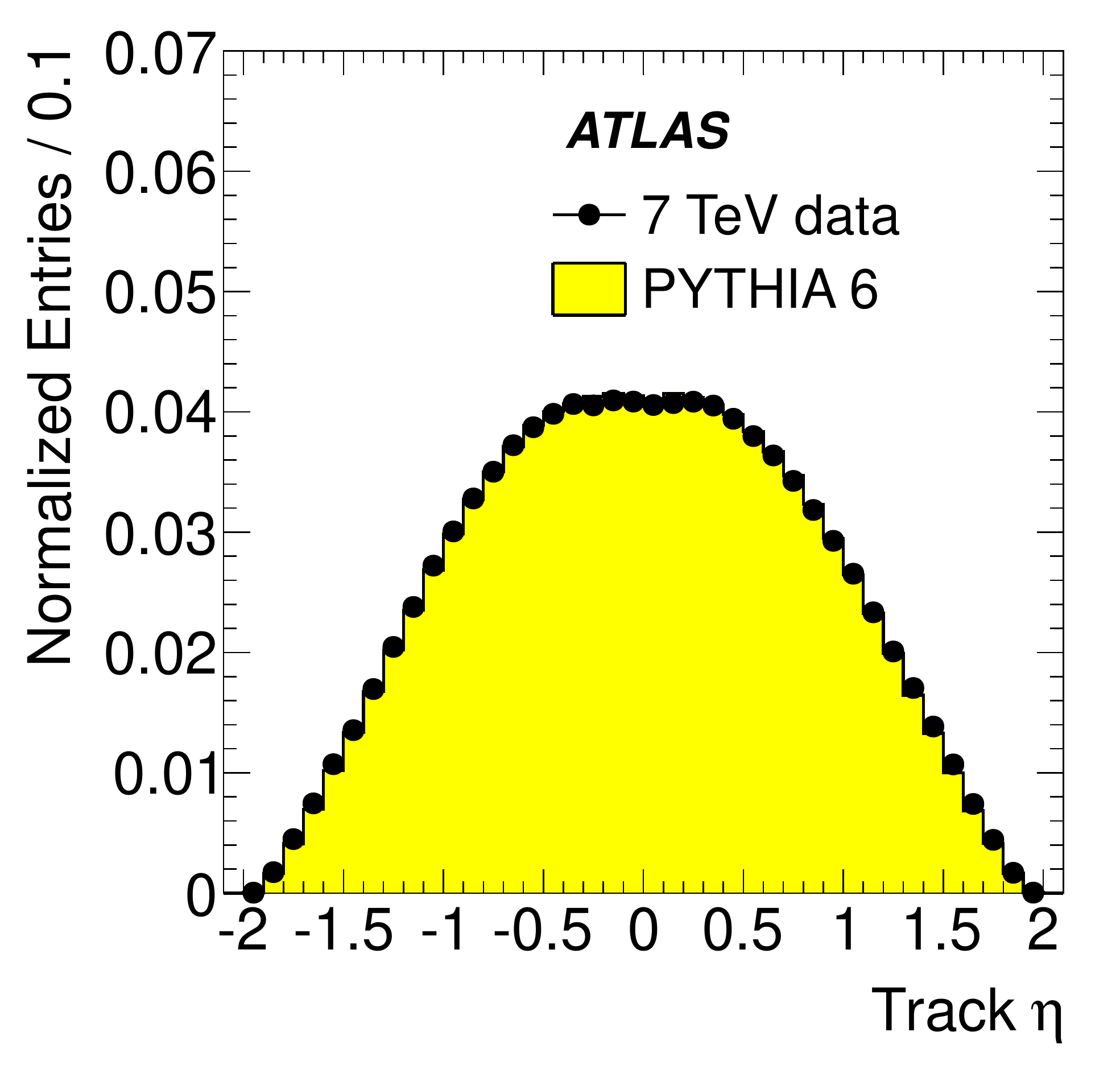}}  
   \subfloat[]{\includegraphics[scale=0.25]{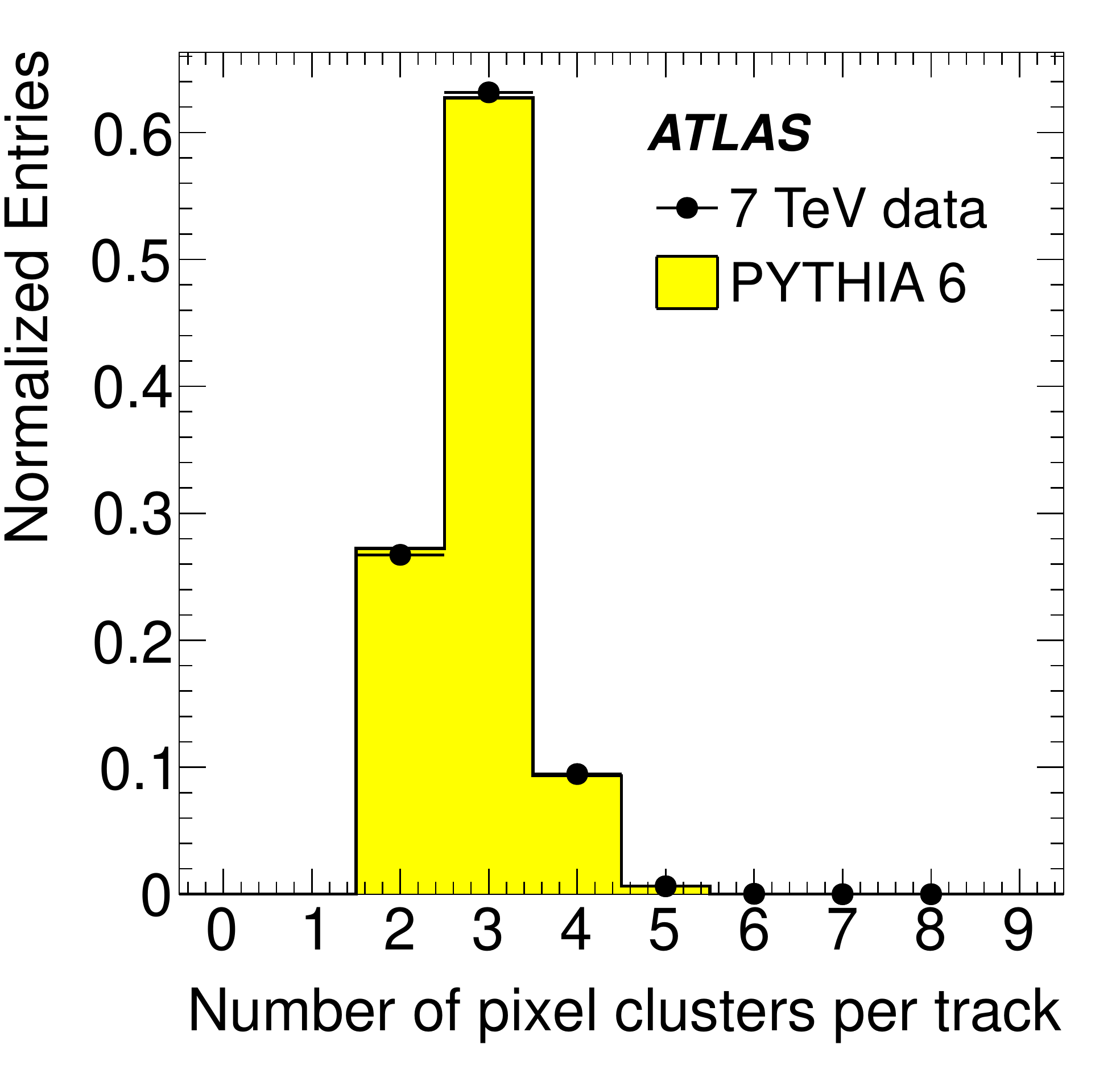}}}
\centerline{
   \subfloat[]{\includegraphics[scale=0.25]{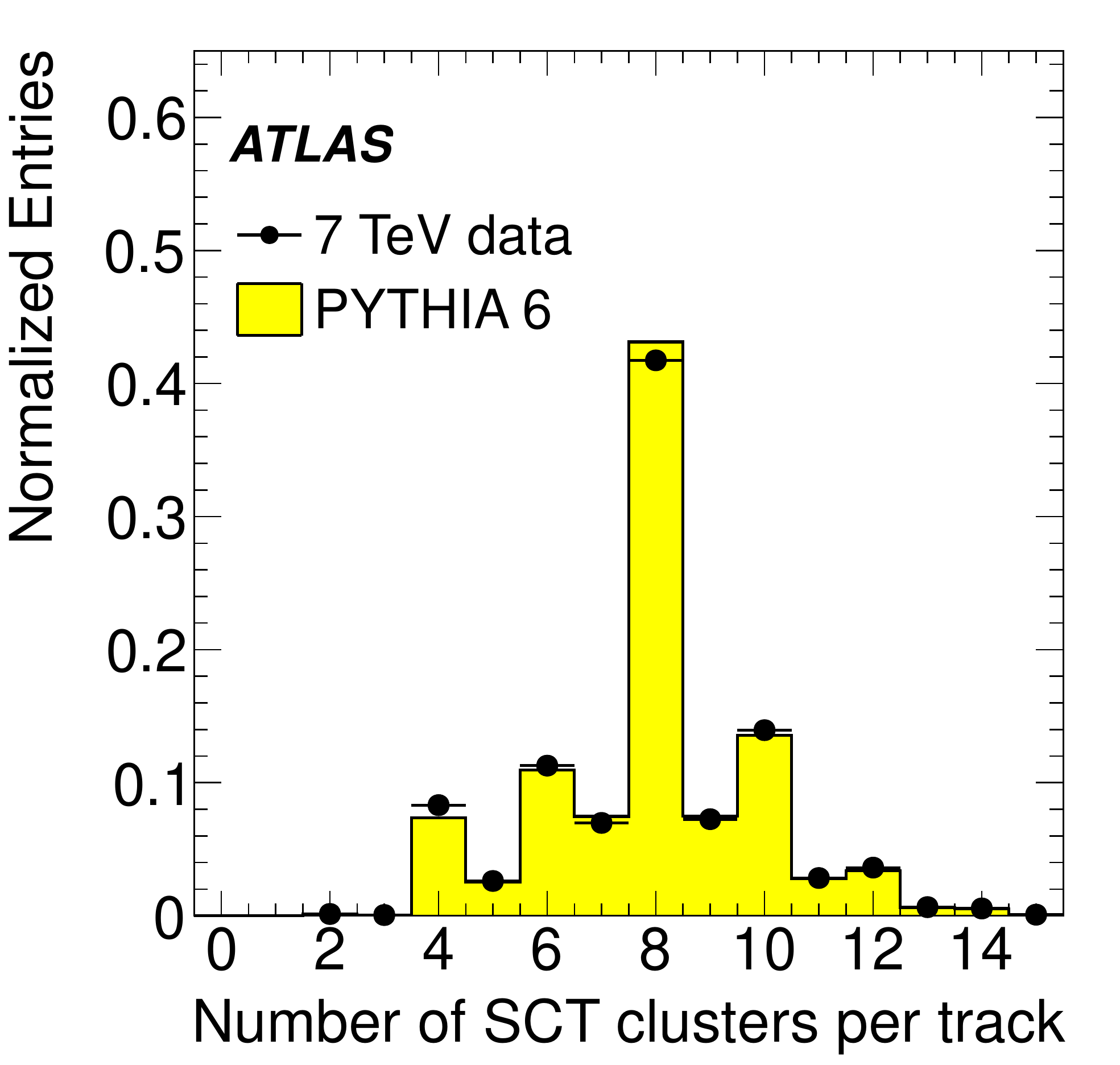}}
   \subfloat[]{\includegraphics[scale=0.25]{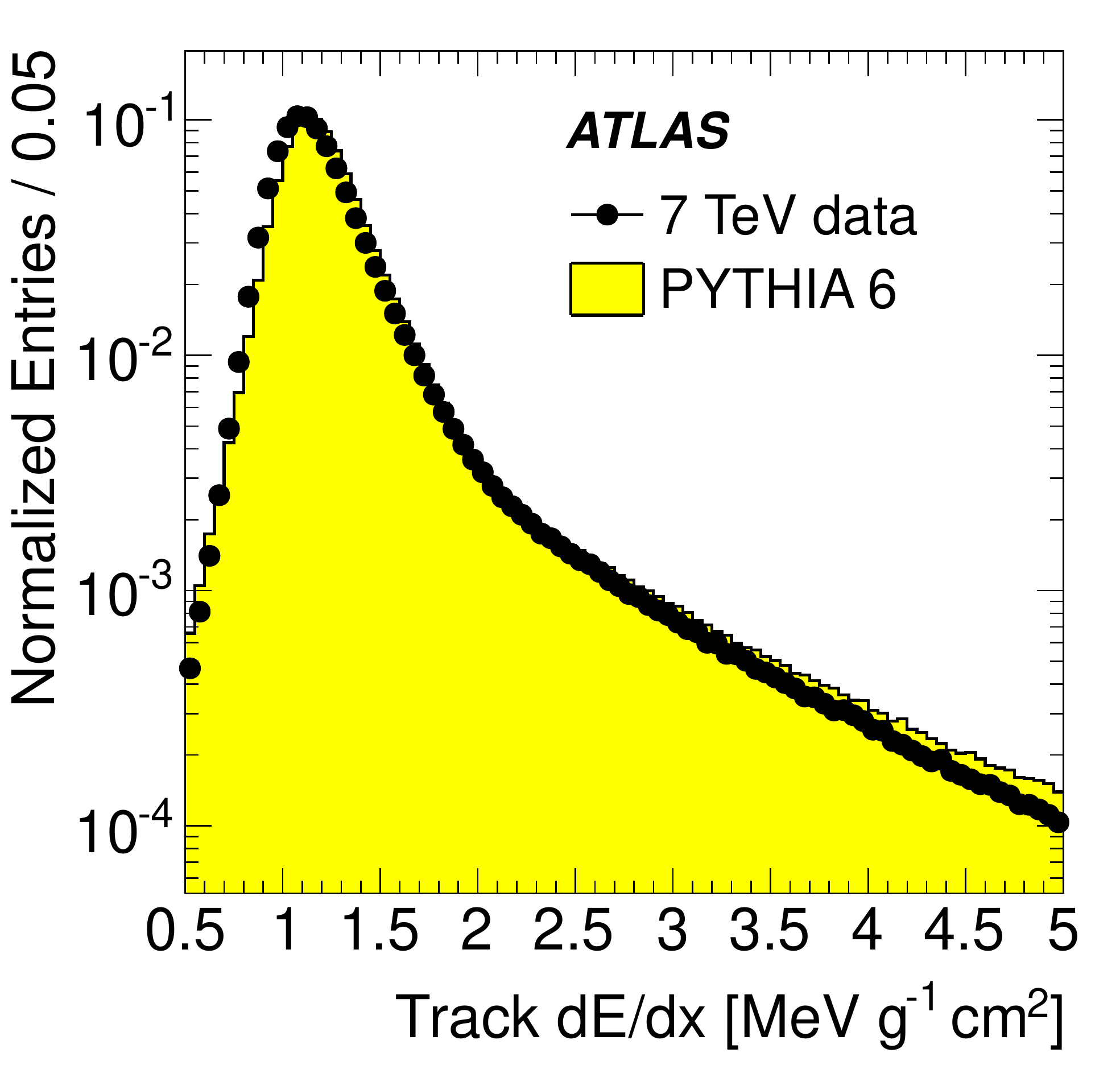}}}
  \caption{Comparison between data (black dots) and MC simulation (histogram) for (a) track $\eta$, (b) number of pixel clusters assigned to the track, (c) number of SCT clusters assigned to the track and (d) the average track energy loss (see text). Statistical uncertainties are smaller than the marker size.}
  \label{fig:trackplots}
\end{figure*}

A comparison between data and MC prediction of track $\eta$, of the number of hits in the pixel and SCT detectors associated with tracks (with a requirement of at least two pixel clusters and two SCT hits) and of average energy loss per track is presented in Fig.~\ref{fig:trackplots}. The distributions agree well, demonstrating a good understanding of track simulation and reconstruction in the inner detector. The slight disagreement in Fig.~\ref{fig:trackplots}(a), where the location of the peak of the average energy loss is overestimated by $\sim$0.05~\MeV\ $\mathrm{g^{-1}}$ $\mathrm{cm^{2}}$ in the MC simulation, is due to the relative abundances of different particle species being slightly different for data and simulation.

The most probable value of the specific energy loss for a pion, kaon or proton hypothesis is para-meterized as a function of the charged particle's Lorentz factor $\beta \gamma$. The measured energy loss is used to calculate the probability $P_{\mathrm{particle}}$ of compatibility with a given hypothesis~\cite{ATLAS-CONF-2011-016}. Kaon candidates are required to satisfy $P_{\mathrm{pion}}<$ 0.1 and $P_{\mathrm{kaon}}>$ 0.84 conditions. The candidate $\phi$(1020) decays are searched for by selecting the oppositely charged track pairs for which both tracks pass the tracking and PID requirements defined above and combine to an invariant mass in the range 1000 $<$ $m(K^+K^-)$ $<$ 1060~\MeV.

\section{Determination of the cross section}\label{sec:defenition}
The fiducial region is divided into eight bins in $\yphi$ and ten bins in $\pTphi$ with bin widths of 0.1 and 70~\MeV, respectively. Unless specifically stated, the cross section is not corrected for the branching fraction of $\phi$(1020)-meson decays to kaons. 

Each $\phi$(1020) candidate is assigned a weight to correct for experimental losses. Firstly, a weight is given for trigger and vertex reconstruction efficiencies~\cite{2010ir}, which have both been measured in data to rapidly increase to 100\% for events with four or more tracks. The trigger and vertex reconstruction efficiencies were found to have a negligible effect on this analysis and were applied on an event-by-event basis. Secondly, a weight is given for track reconstruction and kaon identification efficiencies on a track-by-track basis. These efficiencies are calculated separately for tracks from positively and negatively charged particles, because fewer pixel clusters are expected on the tracks of low-momentum negatively charged particles, which may pass in between two pixel modules due to the tiling and tilt of the modules. The average number of pixel clusters on tracks which pass the selection detailed in Sect.~\ref{sec:event-sel} is $2.96 \pm 0.01$ per positively charged particle and $2.79 \pm 0.01$ per negatively charged particle. Finally, a weight is given on a track-by-track basis to correct for the fraction of selected tracks passing the kinematic selection for which the corresponding truth kaon is outside the kinematic range. Following the determination of the weight of each of the candidate $\phi$(1020), the efficiency-corrected number of reconstructed candidates is determined with a fit to the invariant mass distribution.

The calculation of track reconstruction efficiency, kaon identification efficiency and the subsequent signal yield extraction are explained in the next sections.

\subsection{Track reconstruction efficiency}\label{sec:rec-eff}
The track reconstruction efficiency, $\epsilon_{\mathrm{rec}}$, is based on MC `truth-matching', where generated particles are matched to reconstructed tracks. The method is based on a matching probability evaluated using the number of common hits between particles at generator level and the reconstructed tracks, and is described in Ref.~\cite{2010ir}. The average tracking efficiency for the two tracks of a $\phi \rightarrow K^{+}K^{-}$ decay is about 40\% for the lower $\pTphi$ bins and increases to 65\% in the highest $\pTphi$ bin. It is $\sim$ 50\% for all bins in rapidity.   

To estimate the quality of the MC description of $\epsilon_{\mathrm{rec}}$ in data, the number of tracks passing all cuts in bins of pseudorapidity is divided by the number of tracks passing the cuts with one cut loosened. This efficiency is referred to as the relative efficiency $\epsilon_{\mathrm{rel}}$. The behavior of $\epsilon_{\mathrm{rel}}$ with one fewer pixel cluster or one fewer SCT hit required per track and a lower momentum cut is compared between simulation and data and found to be consistent within 0.5\%. The systematic uncertainty inferred is 0.7\% per track pair.

The dominant source of uncertainty is due to uncertainty in the MC material description, denoted as $\epsilon_{\mathrm{rec}}$(material). It is described in Ref.~\cite{2010ir} and is given in bins of track $\eta$ and $\pT$. The material uncertainty, expressed as a fraction of the corresponding tracking efficiency, is 2--3\% for most tracks accepted in this analysis. To evaluate the impact of this uncertainty, the yield is extracted with the nominal tracking efficiency, and with the nominal tracking efficiency varied up and down by this uncertainty. The systematic uncertainty arising from $\epsilon_{\mathrm{rec}}$(material) is accounted for per bin in $\pTphi$ or $|y_{\phi}|$ and is 5\% per track pair.

The number of reconstructed decays is corrected for the fraction of selected tracks passing the kinematic selection for which the corresponding primary particle is outside the kinematic range. The distributions are subsequently corrected using a MC derived factor to account for the migration of reconstructed $\phi$(1020)-meson candidates into the fiducial volume. The systematic uncertainty arising from this migration correction is evaluated by re-calculating the migration correction after re-wei- ghting the kaon momentum spectrum at particle-level to get a good description of the data at detector level. The variation of the extracted yield using the default and re-weighted migration correction is assigned as a systematic uncertainty and is below 1\%.

The statistical uncertainty on the tracking efficiency, $\epsilon_{\mathrm{rec}}$(stat), is in the range 1--5\% and is propagated as a systematic uncertainty on the cross section. The total systematic uncertainty in the tracking efficiency determination is obtained by adding the previously mentioned components in quadrature and is summarized in Tables \ref{table:result-pT} and \ref{table:result-rapi} as a function of $\pTphi$ and $\yphi$, respectively.

\subsection{Particle identification efficiency}\label{sec:pid-eff}
The particle identification efficiency, $\epsilon_{\mathrm{pid}}$, is extracted from simulation as a function of both $\pK$ and $\eta$. The data sample is not large enough to determine the PID efficiency with a purely data driven technique in bins of $\pK$ and $\eta$. Therefore a data-driven tag-and-probe technique is used to determine the PID in bins of $\pK$ and this is used to rescale the Monte Carlo estimates of the PID efficiency. The data sample is split up into five bins of $\pK$ and the efficiency is measured as the fraction $N_{\mathrm{probe}}/N_{\mathrm{tag}}$, where $N_{\mathrm{probe}}$ is the number of candidates for which both kaons pass the PID requirement of $P_{\mathrm{pion}}<$ 0.1 and $P_{\mathrm{kaon}}>$ 0.84, and $N_{\mathrm{tag}}$ is the number of candidates for which at least the $K^{+}$ or the $K^{-}$ passes. To measure the signal yields $N_{\mathrm{tag}}$ and $N_{\mathrm{probe}}$, the invariant mass distribution in each bin of $\pK$ is fitted with a probability density function (p.d.f.) that describes the signal and background contributions separately and which is detailed in Sect.~\ref{sec:fit-yield}. A final efficiency correction factor is defined by multiplying the two-dimensional efficiency from MC simulation by the ratio of data to MC tag-and-probe efficiencies, which is close to unity for $\pK<$ 500~\MeV, but decreases to a factor of slightly more than 0.3 for 700 $<\pK<$ 800~\MeV. 

The tag-and-probe method is validated using MC simulation by ascertaining that the $\epsilon_{\mathrm{pid}}$ values obtained using MC truth-matching and the tag-and-probe method in bins of $\pTphi$ and $\yphi$ agree within MC statistical uncertainties. The particle identification efficiency decreases with increasing average kaon momentum from $\sim$90\% for 230 $< \pK \leq $ 400~\MeV\ to $\sim$10\% for 700 $<\pK<$ 800~\MeV. 

The systematic uncertainty due to $\epsilon_{\mathrm{pid}}$ is evaluated by fixing the background shape parameters in the tag sample to the values given by the fit to the same-sign background distribution (a maximum uncertainty of 10\%) and by adding the same-sign background samples to the fitted data sets for the tag-and-probe validation in \texttt{\PYS} to vary the signal to background ratio (a maximum uncertainty of 6\%). Possible dependence of the cross section on the choice of $P_{\mathrm{kaon}}$ requirement is tested by varying the requirement by 10\% and is found to be well within the uncertainty due to fixing the background shape parameters. The statistical uncertainty on $\epsilon_{\mathrm{pid}}$ is calculated using a binomial probability distribution, which leads to a relative uncertainty on $\epsilon_{\mathrm{pid}}$ of at most 5\%, denoted by $\epsilon_{\mathrm{pid}}$(stat). These uncertainties (evaluated per bin in $\pTphi$ or $\yphi$) are added in quadrature and are included as systematic uncertainties on the cross section as summarized in Tables~\ref{table:result-pT} and \ref{table:result-rapi}.

\subsection{Signal extraction}\label{sec:fit-yield}
To extract the signal yields, a binned $\chi^2$ fit to the invariant mass spectrum is performed in each region of phase space after applying corrections for the selection efficiencies to the tracks. The signal shape is described by a relativistic Breit--Wigner,
\begin{equation}\label{eq:relBW}
 f_{\mathrm{RBW}}(m;m_{0},\Gamma_{0}) = \frac{m^{2}}{(m^{2}-m_{0}^{2})^{2} + m_{0}^{2} \Gamma^{2}(m) }, 
\end{equation}
where the mass-dependent width is given by
\begin{equation}\label{eq:Gamma_m_phi}
 \Gamma(m) = \Gamma_{0} \left[ \frac{m^{2}-4m_{K}^{2}}{m_{0}^{2}-4m_{K}^{2}} \right]^{3/2}. 
\end{equation}
In Eq. (\ref{eq:relBW}), $m_{0}$ is fixed to the $\phi(1020)$-meson mass of 1019.45~\MeV\ \cite{pdg}, $\Gamma_{0}$ to the natural width of 4.26~\MeV\ \cite{pdg}, and $m_{K}$ in Eq.~(\ref{eq:Gamma_m_phi}) is the charged kaon mass~\cite{pdg}. 

The signal shape is convolved with a Gaussian resolution function, with the mean and standard deviation left free in the fit. The mean of the Gaussian is interpreted as the actual value of the $\phi$(1020) mass, while its standard deviation corresponds to the experimental resolution. The values obtained from the fits are in the range $\sigma_{\mathrm{exp}} = $1.0--2.5~\MeV. 

This signal description is added to an empirical background description,
\begin{eqnarray}\label{bkg-pdf}
&& f_{\mathrm{BKG}}(m) = \left(1- e^{(2m_{K}-m)/C}\right) \cdot
\left(\frac{m}{2m_{K}}\right)^{A} \nonumber \\
&&+B\left( \frac{m}{2m_{K}}-1\right), \nonumber \\
\end{eqnarray}
where $A$, $B$ and $C$ determine the background shape. Initial values for $A$, $B$ and $C$ are found by fitting the background p.d.f. to a sample of events with two kaons of the same charge. This same-sign sample contains the same sources of combinatorial background as the nominal selection but no true $\phi$(1020) mesons, and so it provides a good initial description of the background shape. It was checked that the background model provides stable fitting results in all bins in $\pTphi$ and $|y_{\phi}|$ for the same-sign sample. 

\begin{figure*}
  \centerline{
\subfloat[]{\includegraphics[scale=0.4]{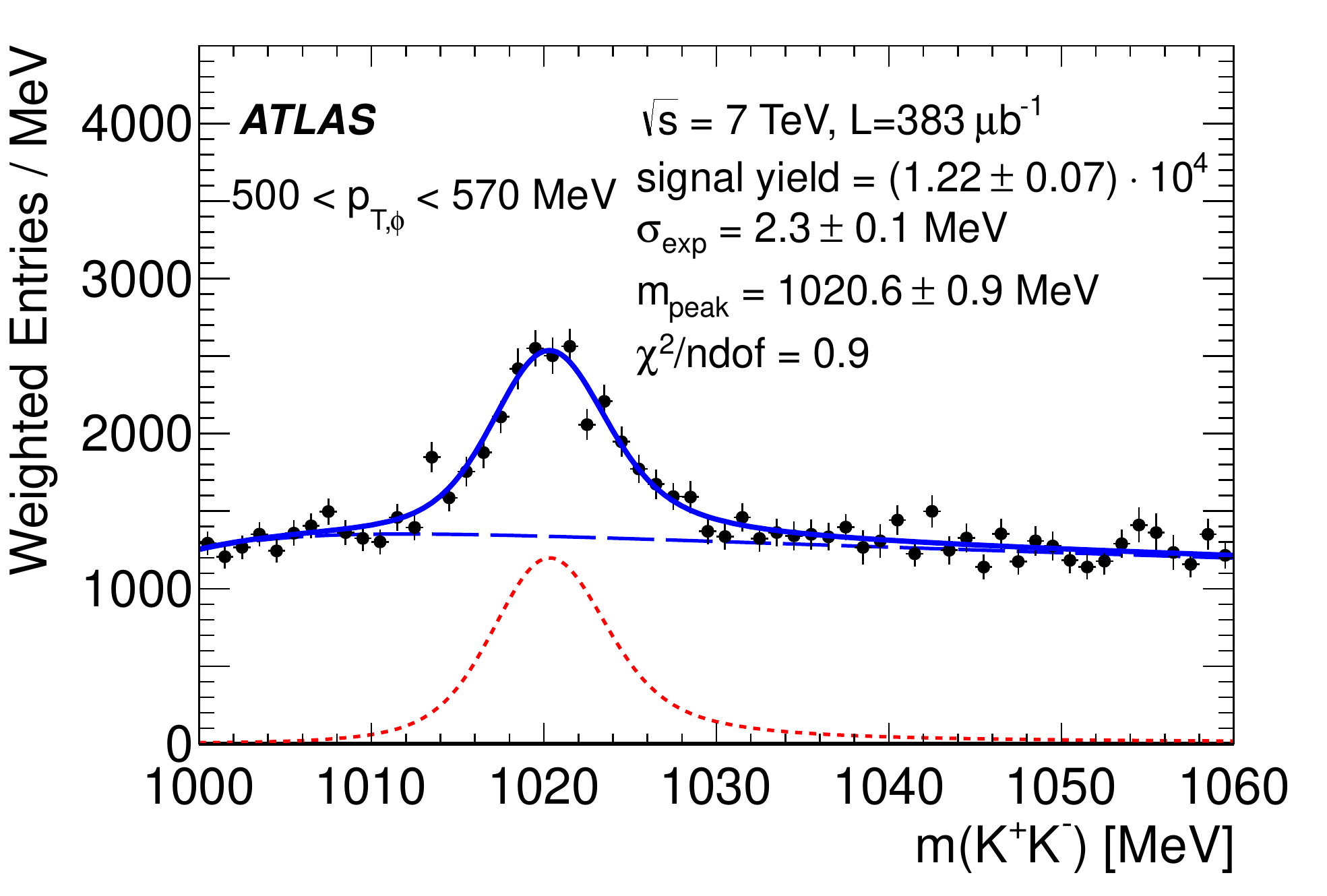}}
\subfloat[]{\includegraphics[scale=0.4]{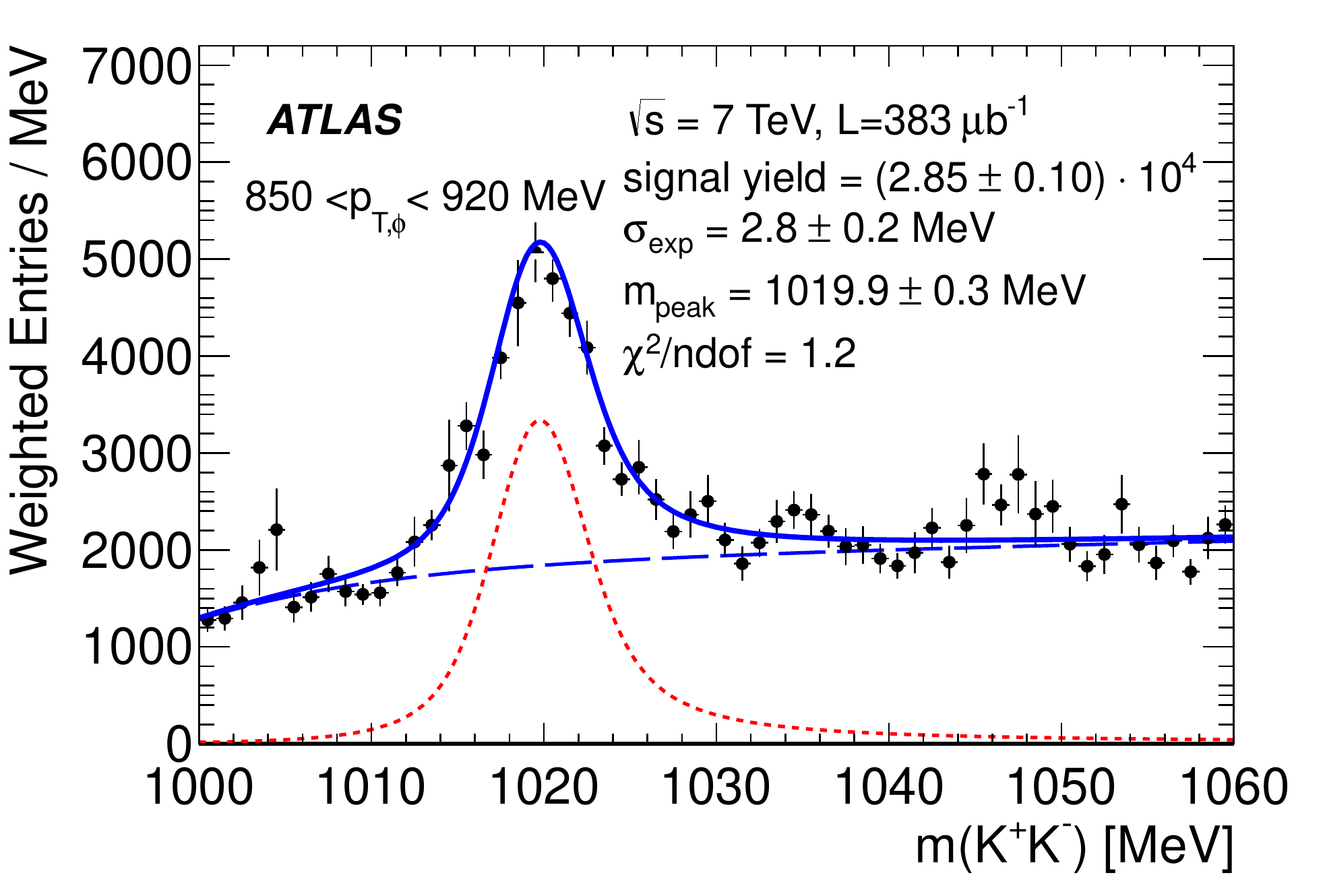}}}

\centerline{
\subfloat[]{\includegraphics[scale=0.4]{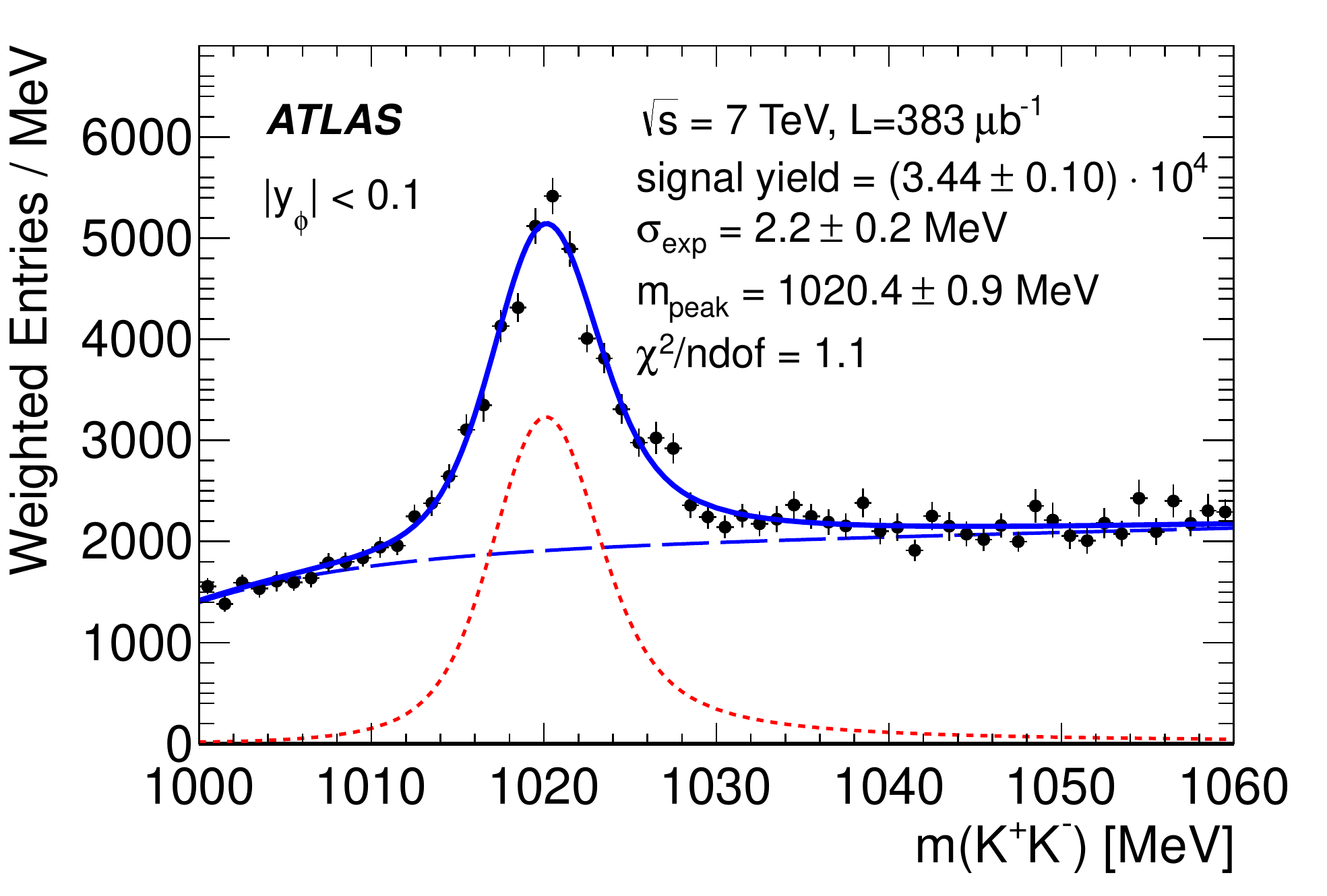}}
 \subfloat[]{\includegraphics[scale=0.4]{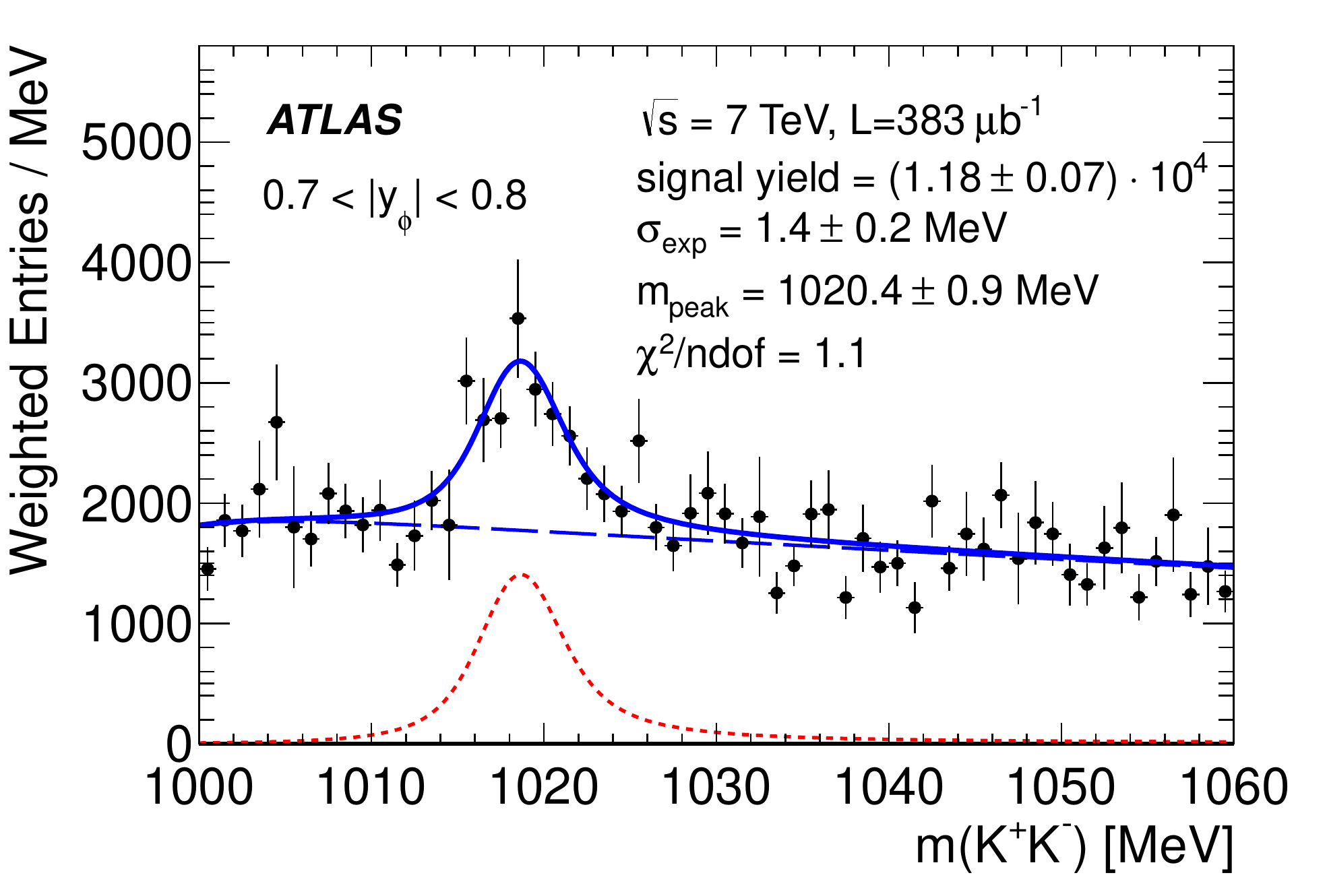}}}

 \caption{Examples of invariant $K^{+}K^{-}$ mass distributions in the data (dots) compared to results of the fits (solid lines), as described in the text, for (a) the lowest $\pTphi$ bin, (b) one of the middle $\pTphi$ bins, (c) the most central $\yphi$ bin and (d) most forward $\yphi$ bin. The dashed curves show the background contribution and the dotted red curves demonstrates the signal contributions, with paremeters listed in the legend.}
 \label{fig:invmassfit}
\end{figure*}

Fits of the invariant mass of $K^{+}K^{-}$ pairs are shown in Fig.~\ref{fig:invmassfit} for four regions. It was found that the maximum of the signal peak, $m_{\mathrm{peak}}$, is shifted upwards by almost 1~\MeV\ for the lowest $\pTphi$ bin. This is covered by the uncertainty on the momentum scale for the low-momentum tracks.

Three tests are conducted to estimate the systematic uncertainty on the extracted signal yield due to uncertainty on the signal, background shape and detector resolution. Firstly, the signal is extracted using a non-relativistic Breit--Wigner lineshape convolved with a Gaussian to describe the signal shape. This leads to a conservative estimate of the uncertainties in the extracted signal of 5--6\%, which are evaluated bin-by-bin in $\pTphi$ and $\yphi$. Secondly, the extracted yield changes by at most 2\% if the signal shape is convolved with a Crystal Ball~\cite{thesis-crystal-ball} resolution function, rather than a Gaussian. Thirdly, the extracted yields vary by at most 3\% if the background p.d.f. is fitted to the sample of same-sign pairs of tracks in each bin and the shape is fixed to the result of this fit. Adding the relative changes in the yield in quadrature, a conservative estimate of 6--7\% is assigned to the systematic uncertainty and summarized in Tables~\ref{table:result-pT} and \ref{table:result-rapi}. 

The cross section $\sigma^{\mathnormal{i}}_{\mathrm{bin}}$ in bin $\mathnormal{i}$ is determined by
\begin{equation}\label{eq:cross}
\sigma^{\mathnormal{i}}_{\mathrm{bin}} = \frac{N_{\mathnormal{i}} } {\mathcal{L}},
\end{equation}
where $\mathcal{L}$ is the integrated luminosity and $N_{\mathnormal{i}}$ is the number of efficiency-corrected reconstructed $\phi \rightarrow K^{+}K^{-}$ candidates in bin $\mathnormal{i}$. 

\begin{figure*}
  \centering
\includegraphics[width=0.47\textwidth]{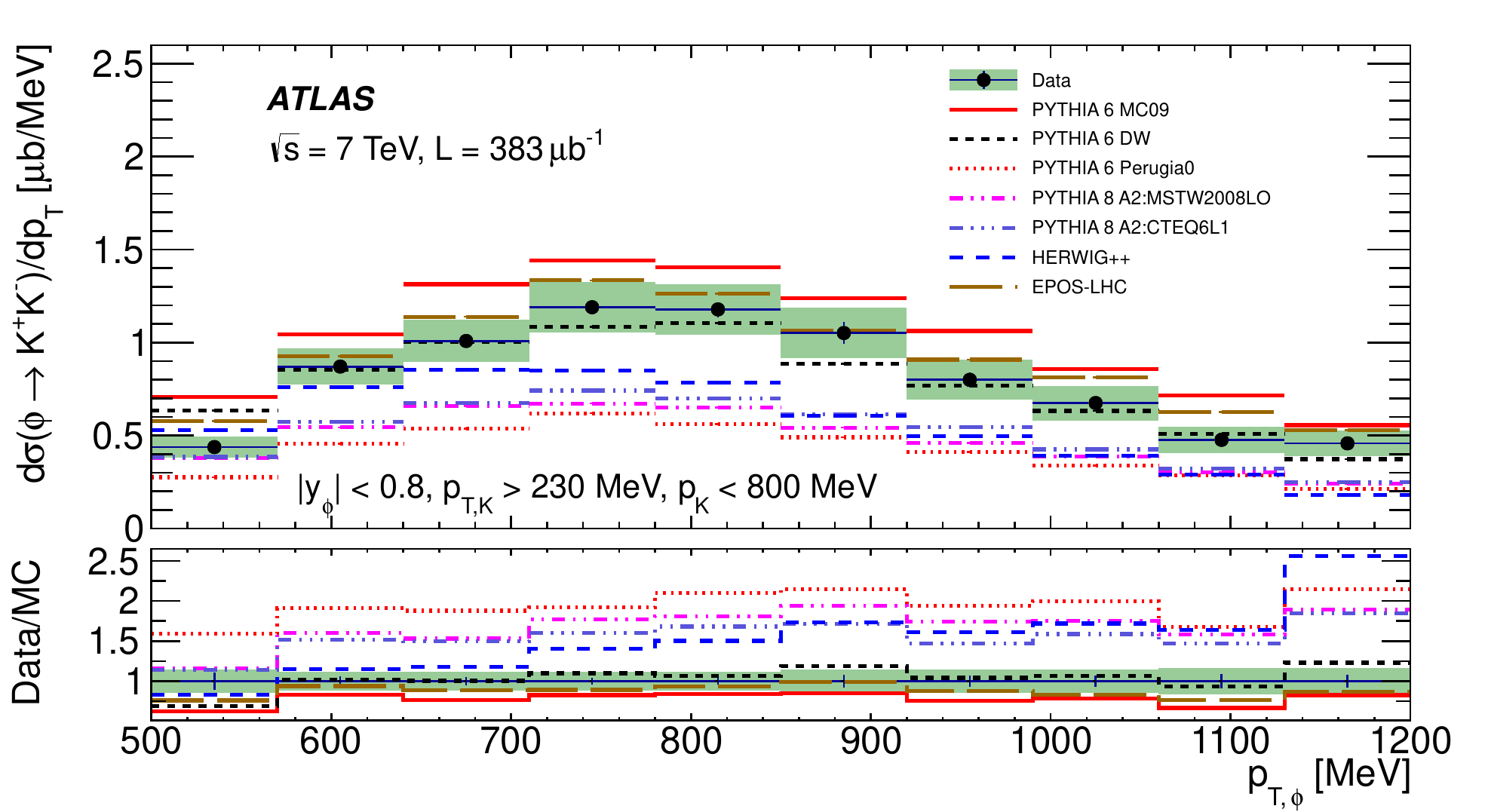} 
\includegraphics[width=0.47\textwidth]{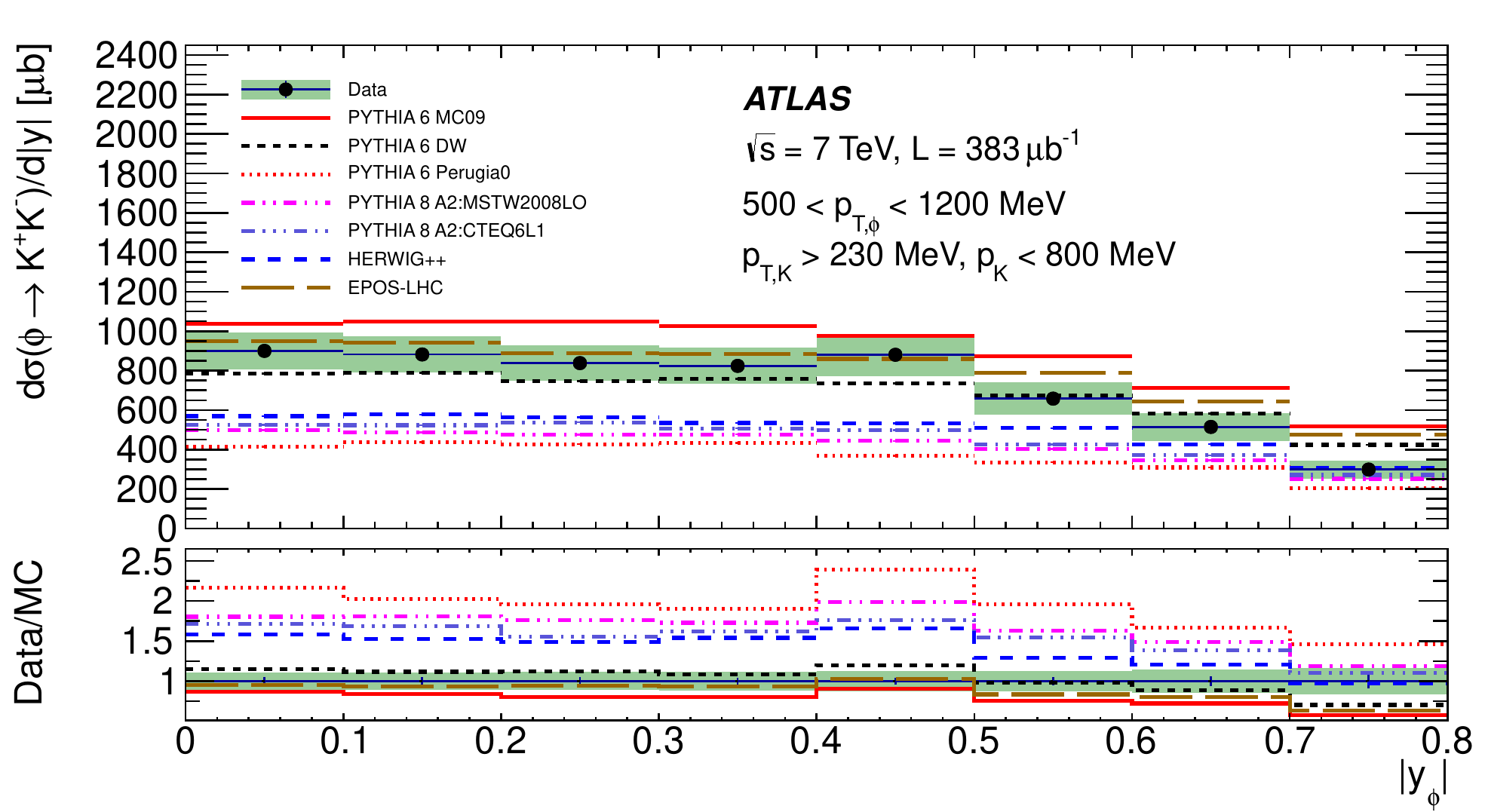}
\caption{The $\phi\mathrm{(1020)} \rightarrow K^{+} K^{-} $ cross section in the fiducial region, with 500 $< \pTphi <$ 1200~\MeV, $|y_{\phi}|<$ 0.8, $\pTK>$ 230~\MeV\ and kaon momentum $\pK<$ 800~\MeV, as a function of $ \pTphi $  (left) and $|y_{\phi}|$ (right). The error bars represent the statistical uncertainty and the green boxes represent the quadratic sum of the statistical and systematic uncertainties. The 3.5\% uncertainty on the luminosity is not included. The data are compared to various MC expectations as described in the legends.}
  \label{fig:cross-section-probe}
\end{figure*}

\begin{table*}
\centering
 \caption{The fitted number of $\phi$(1020) candidates (Signal), the differential production cross section $\mathrm{d}\sigma/\mathrm{d} \pT$ [$\mu$b/\MeV] of $\phi \rightarrow K^{+} K^{-}$ and its statistical uncertainty in bins of $\pTphi$ with 500~$<\pTphi<$~1200~\MeV, $\yphi<$~0.8, $\pTK>$~230~\MeV\ and $\pK<$~800~\MeV\ and the systematic uncertainties due to track reconstruction efficiency ($\epsilon_{\mathrm{rec}}$), kaon identification ($\epsilon_{\mathrm{pid}}$) and fitting procedure. The uncertainty on the luminosity is 3.5\%.}
\begin{tabular}{l c c c c c}
\hline
 & & & \multicolumn{3}{c}{Systematic uncertainty} [$\mu$b/\MeV] \\ 
\noalign{\smallskip}\hline\noalign{\smallskip}
    Bin [\MeV]& Signal [in units of $10^{4}$] & $\mathrm{d}\sigma/\mathrm{d}\pT$ [$\mu$b/\MeV] & $\epsilon_{\mathrm{rec}}$ & $\epsilon_{\mathrm{pid}}$ & Fitting \\
\noalign{\smallskip}\hline\noalign{\smallskip}   
$500<\pTphi \leq 570$ &1.22 $\pm$ 0.07 & 0.44 $\pm$ 0.03 & $\pm$ 0.03  & $\pm$ 0.03  & $\pm$ 0.07 \\
$570< \pTphi \leq 640$ &2.34 $\pm$ 0.09 & 0.87 $\pm$ 0.04 & $\pm$ 0.06  & $\pm$ 0.05  & $\pm$ 0.14 \\
$640<\pTphi \leq 710$ &2.71 $\pm$ 0.10 & 1.01 $\pm$ 0.04 & $\pm$ 0.06  & $\pm$ 0.06  & $\pm$ 0.16 \\
$710<\pTphi \leq 780$ &3.19 $\pm$ 0.11 & 1.19 $\pm$ 0.04 & $\pm$ 0.07  & $\pm$ 0.09  & $\pm$ 0.19 \\
$780<\pTphi \leq 850$ &3.16 $\pm$ 0.11 & 1.18 $\pm$ 0.04 & $\pm$ 0.06  & $\pm$ 0.10  & $\pm$ 0.19 \\
$850<\pTphi \leq 920$ &2.85 $\pm$ 0.10 & 1.05 $\pm$ 0.04 & $\pm$ 0.05  & $\pm$ 0.09  & $\pm$ 0.17 \\
$920<\pTphi \leq 990$ &2.15 $\pm$ 0.09 & 0.79 $\pm$ 0.04 & $\pm$ 0.03  & $\pm$ 0.08  & $\pm$ 0.15 \\
$990<\pTphi \leq 1060$ &1.81 $\pm$ 0.07 & 0.67 $\pm$ 0.04 & $\pm$ 0.03  & $\pm$ 0.07  & $\pm$ 0.13 \\
$1060<\pTphi \leq 1130$ &1.30 $\pm$ 0.06 & 0.48 $\pm$ 0.04 & $\pm$ 0.02  & $\pm$ 0.05  & $\pm$ 0.09 \\
$1130<\pTphi \leq 1200$ &1.23 $\pm$ 0.08 & 0.46 $\pm$ 0.04 & $\pm$ 0.02  & $\pm$ 0.06  & $\pm$ 0.09 \\
    \hline
\end{tabular} 
     \label{table:result-pT}
 \end{table*}

\begin{table*}
\centering
 \caption{The fitted number of $\phi$(1020) candidates (Signal), the differential production cross section $\mathrm{d}\sigma/\mathrm{d}|y|$ [mb] of $\phi \rightarrow K^{+} K^{-} $ and its statistical uncertainty in bins of $|y_{\phi}|$ with 500~$<\pTphi<$~1200~\MeV, $\yphi<$~0.8, $\pTK>$~230~\MeV\ and $\pK<$~800~\MeV\ and the systematic uncertainties due to track reconstruction efficiency ($\epsilon_{\mathrm{rec}}$), kaon identification ($\epsilon_{\mathrm{pid}}$) and fitting procedure. The uncertainty on the luminosity is 3.5\%.}
\begin{tabular}{l c c c c c}
\hline
& & & \multicolumn{3}{c}{Systematic uncertainty} [mb]\\ 
\noalign{\smallskip}\hline\noalign{\smallskip}
 Bin & Signal [in units of $10^{4}$] & $\mathrm{d}\sigma/\mathrm{d}|y|$ [mb] & $\epsilon_{\mathrm{rec}}$ & $\epsilon_{\mathrm{pid}}$ & Fitting \\
\noalign{\smallskip}\hline\noalign{\smallskip}   
$0.0<|y_{\mathrm{\phi}}| \leq 0.1$&3.44 $\pm$ 0.10 & 0.90 $\pm$ 0.03 & $\pm$ 0.04  & $\pm$ 0.06  & $\pm$ 0.21 \\
$0.1<|y_{\mathrm{\phi}}| \leq 0.2$&3.39 $\pm$ 0.10 & 0.88 $\pm$ 0.03 & $\pm$ 0.04  & $\pm$ 0.07  & $\pm$ 0.20 \\
$0.2<|y_{\mathrm{\phi}}| \leq 0.3$&3.22 $\pm$ 0.09 & 0.84 $\pm$ 0.03 & $\pm$ 0.04  & $\pm$ 0.06  & $\pm$ 0.19 \\
$0.3<|y_{\mathrm{\phi}}| \leq 0.4$&3.18 $\pm$ 0.09 & 0.82 $\pm$ 0.03 & $\pm$ 0.04  & $\pm$ 0.06  & $\pm$ 0.19 \\
$0.4<|y_{\mathrm{\phi}}| \leq 0.5$&3.36 $\pm$ 0.11 & 0.88 $\pm$ 0.03 & $\pm$ 0.05  & $\pm$ 0.08  & $\pm$ 0.20 \\
$0.5<|y_{\mathrm{\phi}}| \leq 0.6$&2.53 $\pm$ 0.12 & 0.66 $\pm$ 0.03 & $\pm$ 0.04  & $\pm$ 0.06  & $\pm$ 0.18 \\
$0.6<|y_{\mathrm{\phi}}| \leq 0.7$&2.01 $\pm$ 0.11 & 0.51 $\pm$ 0.02 & $\pm$ 0.03  & $\pm$ 0.05  & $\pm$ 0.14 \\
$0.7<|y_{\mathrm{\phi}}| \leq 0.8$&1.18 $\pm$ 0.07 & 0.30 $\pm$ 0.02 & $\pm$ 0.02  & $\pm$ 0.04  & $\pm$ 0.08 \\

\noalign{\smallskip}\hline	
\end{tabular} 
   \label{table:result-rapi}
 \end{table*}

\section{Results}\label{sec:result}
The differential $\phiKK$ cross section in the fiducial region 500 $< \pTphi <$ 1200~\MeV, $|y_{\phi}|<$ 0.8, kaon transverse momentum $\pTK>$ 230~\MeV\ and kaon momentum $\pK<$ 800~\MeV\ is shown in Fig.~\ref{fig:cross-section-probe} as functions of $\pTphi$ and $|y_{\phi}|$ and compared to simulation. Tables \ref{table:result-pT} and \ref{table:result-rapi} give the differential cross sections and the relevant systematic uncertainties. The total statistical uncertainty ranges from 3\% to 8\% and the total systematic uncertainty is 8--12\%. The uncertainty on the luminosity is 3.5\%~\cite{Aad:2013ucp} for all bins. The integrated cross section is calculated as the sum of the differential cross sections as a function of $\pTphi$. This determination is less sensitive to mismodelling of the $\pTphi$ distribution and is measured to be $\sigma_{\phiKK}$ = 570 $\pm$ 8 (stat) $\pm$ 68 (syst) $\pm$ 20 (lumi)~$\mathrm{\mu b}$.

The fiducial cross section increases as a function of $\pTphi$ in the range 500--700~\MeV, reaches a maximum at $\pTphi \sim$ 750~\MeV\ and decreases for $\pTphi \geq$ 850~\MeV. The increase in the number of measured decays as $\pTphi$ rises to 700~\MeV\ is due to the cut on kaon transverse momentum $\pTK >$ 230~\MeV. Being within the rapidity plateau at the LHC given that $\yphi < 0.8$, the total $\phi$(1020) cross section differential in $\yphi$ is expected to be constant. The fiducial cross section is however seen to decrease from $\yphi \geq$ 0.5. This is due to the $\pK <$ 800~\MeV\ requirement for efficient PID which excludes an increasing fraction of kaons as the rapidity increases. 

The cross section is best described by the \texttt{PY- THIA 6} tune DW and by the \texttt{EPOS-LHC} tune. These provide a good description for the $\pTphi$ and $\yphi$ dependencies as well as for the total yield. The \texttt{\PYS} MC09 tune slightly overestimates the data in the fiducial region. The \texttt{\PYS} Perugia0 tune underestimates the cross section by around a factor of two compared to the data in the whole fiducial volume. The two \texttt{\PYE} A2 tunes, based on different PDFs, show similar predictions for the cross section, which are also about a factor of two too small. \texttt{Herwig++} provides a good description for the cross section for $\pTphi < 700$~\MeV\ and for $\yphi>0.6$, but exhibits an overly steeply falling $\pTphi$ dependence, such that the cross section is underestimated for $\pTphi > 700$~\MeV\ and in the mid-rapidity range $\yphi<0.6$.

\section{Extrapolated cross section}\label{sec:discussion}
The kaon momenta requirements arising from tracking and PID cuts ($\pTK>$ 230~\MeV\ and $\pK<$ 800~\MeV) reject a significant number of $\phi \rightarrow K^{+} K^{-}$ candidates. In order to allow comparison with other measurements, the cross section in the fiducial region is extrapolated to a cross section in the kinematic region 500 $<\pTphi<$ 1200~\MeV\ and central rapidity $\yphi<$ 0.5, using MC particle level information. The variation of the expected correction between the different generators considered is 10\% and is included as an additional systematic uncertainty on the extrapolated result. A correction for the branching fraction is also applied. The systematic uncertainty on the branching fraction is 1\%~\cite{pdg}. The extrapolation is done with \texttt{\PYS}, because the cross section's dependence on $\pTphi$ within the fiducial region is well described by this generator, as shown in Fig.~\ref{fig:cross-section-probe}. The extrapolation is restricted to $\yphi<$ 0.5, where the fiducial acceptance is large, over 70\%. The extrapolation factor is 2.78 for the lowest $\pTphi$ bin, then decreases to 1.08 at $\pTphi \sim$ 900~\MeV\ and becomes 1.21 in highest $\pTphi$ bin.

\begin{figure}
  \centering
\includegraphics[width=0.49\textwidth]{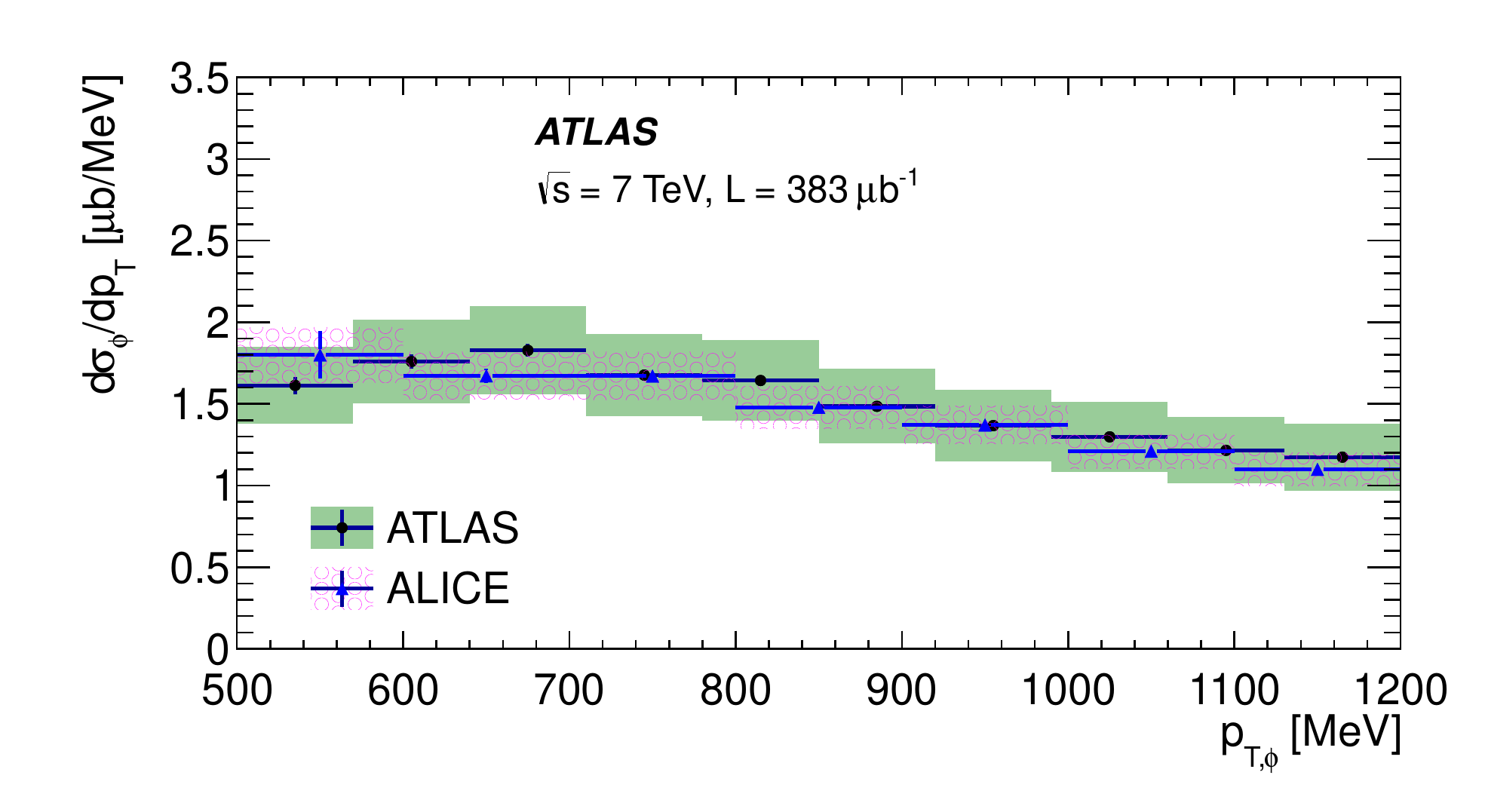}
\caption{The $\phi$(1020)-meson cross section as a function of $\pTphi$, extrapolated using \texttt{\PYS} to the kinematic region with 500 $< \pTphi <$ 1200~\MeV\ and $|y_{\phi}|<$ 0.5, is compared to the measurement by the ALICE Collaboration~\cite{Abelev:2012hy}. The error bars represent the statistical uncertainty and the boxes represent the quadratic sum of the statistical and systematic uncertainties. The 3.5\% uncertainty on the luminosity is not included.}
  \label{fig:cross-section-ATLAS-ALICE}
\end{figure}

The extrapolated cross section is compared to the measurement by the ALICE Collaboration of the $\phi$(1020) production cross section as described in Ref.~\cite{Abelev:2012hy}. A comparison between the cross section measurements is shown in Fig.~\ref{fig:cross-section-ATLAS-ALICE}. The measurements as a function of $\pTphi$ are in agreement to within 10\% in the first two bins and to within 3\% in the other bins, which is well within the systematic uncertainties.

\section{Summary}\label{sec:summary}
This paper presents a measurement of the differential production cross section of the $\phi$(1020) meson using the $K^{+}K^{-}$ decay mode and 383 $\mathrm{\mu b ^{-1}}$ of 7~\TeV\ $pp$ collision data collected with the ATLAS experiment at the LHC. To avoid model-dependent extrapolations outside the detector acceptance, the cross section is measured in a fiducial region, with 500 $< \pTphi <$ 1200~\MeV, $|y_{\phi}|<$ 0.8, kaon $\pTK>$ 230~\MeV\ and kaon momentum $\pK<$ 800~\MeV\ requirements, which are determined by particle identification and track reconstruction constraints.

The $\phi$(1020) production cross section is in agreement with the predictions of the MC generator tunes \texttt{EPOS-LHC} and \texttt{\PYS} DW. \texttt{\PYS} predictions using different tunes are observed to differ significantly. The cross section is also underestimated by \texttt{\PYE} and by \texttt{Herwig++}. This measurement can provide useful input for tuning and development of phenomenological models in order to improve MC generators.

\section{Acknowledgements}
We thank CERN for the very successful operation of the LHC, as well as the
support staff from our institutions without whom ATLAS could not be
operated efficiently.

We acknowledge the support of ANPCyT, Argentina; YerPhI, Armenia; ARC,
Australia; BMWF and FWF, Austria; ANAS, Azerbaijan; SSTC, Belarus; CNPq and FAPESP,
Brazil; NSERC, NRC and CFI, Canada; CERN; CONICYT, Chile; CAS, MOST and NSFC,
China; COLCIENCIAS, Colombia; MSMT CR, MPO CR and VSC CR, Czech Republic;
DNRF, DNSRC and Lundbeck Foundation, Denmark; EPLANET, ERC and NSRF, European Union;
IN2P3-CNRS, CEA-DSM/IRFU,\\France; GNSF, Georgia; BMBF, DFG, HGF, MPG and AvH
Foundation, Germany; GSRT and NSRF,\\Greece; ISF, MINERVA, GIF, DIP and Benoziyo Center,
Israel; INFN, Italy; MEXT and JSPS, Japan;\\CNRST, Morocco; FOM and NWO,
Netherlands; BRF and RCN, Norway; MNiSW, Poland; GRICES and FCT, Portugal; MNE/IFA, Romania; MES of Russia and ROSATOM, Russian Federation; JINR; MSTD,
Serbia; MSSR, Slovakia; ARRS and MIZ\v{S}, Slovenia; DST/NRF, South Africa;
MICINN, Spain; SRC and Wallenberg Foundation, Sweden; SER, SNSF and Cantons of
Bern and Geneva, Switzerland; NSC, Taiwan; TAEK, Turkey; STFC, the Royal
Society and Leverhulme Trust, United Kingdom; DOE and NSF, United States of
America.

The crucial computing support from all WLCG partners is acknowledged
gratefully, in particular from CERN and the ATLAS Tier-1 facilities at
TRIUMF (Canada), NDGF (Denmark, Norway, Sweden), CC-IN2P3 (France),
KIT/GridKA (Germany), INFN-CNAF (Italy), NL-T1 (Netherlands), PIC\\(Spain),
ASGC (Taiwan), RAL (UK) and BNL\\(USA) and in the Tier-2 facilities
worldwide.

\clearpage
\bibliographystyle{atlasBibStyleWoTitle}
\bibliography{references}
\onecolumn
\clearpage
\begin{flushleft}
{\Large The ATLAS Collaboration}

\bigskip

G.~Aad$^{\rm 48}$,
T.~Abajyan$^{\rm 21}$,
B.~Abbott$^{\rm 111}$,
J.~Abdallah$^{\rm 12}$,
S.~Abdel~Khalek$^{\rm 115}$,
A.A.~Abdelalim$^{\rm 49}$,
O.~Abdinov$^{\rm 11}$,
R.~Aben$^{\rm 105}$,
B.~Abi$^{\rm 112}$,
M.~Abolins$^{\rm 88}$
% ...
\bigskip
\\
% ...
$^{11}$ Institute of Physics, Azerbaijan Academy of Sciences, Baku, Azerbaijan\\
$^{12}$ Institut de F{\'\i}sica d'Altes Energies and Departament de F{\'\i}sica de la Universitat Aut{\`o}noma de Barcelona and ICREA, Barcelona, Spain\\
% ...
$^{21}$ Physikalisches Institut, University of Bonn, Bonn, Germany\\
% ...
$^{48}$ Fakult{\"a}t f{\"u}r Mathematik und Physik, Albert-Ludwigs-Universit{\"a}t, Freiburg, Germany\\
$^{49}$ Section de Physique, Universit{\'e} de Gen{\`e}ve, Geneva, Switzerland\\
% ...
$^{88}$ Department of Physics and Astronomy, Michigan State University, East Lansing MI, United States of America\\
% ...
$^{105}$ Nikhef National Institute for Subatomic Physics and University of Amsterdam, Amsterdam, Netherlands\\
% ...
$^{111}$ Homer L. Dodge Department of Physics and Astronomy, University of Oklahoma, Norman OK, United States of America\\
$^{112}$ Department of Physics, Oklahoma State University, Stillwater OK, United States of America\\
% ...
$^{115}$ LAL, Universit{\'e} Paris-Sud and CNRS/IN2P3, Orsay, France\\
% ...
\end{flushleft}


\end{document}